\documentclass{article}
\usepackage{ijcai23}

\usepackage{times}
\usepackage{soul}
\usepackage{url}
\usepackage[hidelinks]{hyperref}
\usepackage[utf8]{inputenc}
\usepackage[font={bf}]{caption}
\usepackage{graphicx}
\usepackage{amsmath}
\usepackage{amsthm}
\usepackage{booktabs}
\usepackage{algorithm}
\usepackage{algorithmic}
\usepackage[switch]{lineno}

\usepackage{amsmath,amssymb,amsfonts}
\usepackage{makecell,multirow,diagbox}

\usepackage{subfig}

\title{GraphPub: Generation of Differential Privacy Graph with High Availability}

\author{
Wanghan Xu$^1$\and
Bin Shi$^{1*}$\and
Ao Liu$^1$\and
Jiqiang Zhang$^1$\and
Bo Dong$^1$
\affiliations
$^1$Xi’an Jiaotong University
\emails
WhoisYt@stu.xjtu.edu.cn
}

\begin{document}

\maketitle

\begin{abstract}
    In recent years, with the rapid development of graph neural networks (GNN), more and more graph datasets have been published for GNN tasks. However, when an upstream data owner publishes graph data, there are often many privacy concerns, because many real-world graph data contain sensitive information like person's friend list. Differential privacy (DP) is a common method to protect privacy, but due to the complex topological structure of graph data, applying DP on graphs often affects the message passing and aggregation of GNN models, leading to a decrease in model accuracy. In this paper, we propose a novel graph edge protection framework, graph publisher (GraphPub), which can protect graph topology while ensuring that the availability of data is basically unchanged. Through reverse learning and the encoder-decoder mechanism, we search for some false edges that do not have a large negative impact on the aggregation of node features, and use them to replace some real edges. The modified graph will be published, which is difficult to distinguish between real and false data. Sufficient experiments prove that our framework achieves model accuracy close to the original graph with an extremely low privacy budget.
\end{abstract}

\section{Introduction}

Graph neural networks (GNN)~\cite{wu2020comprehensive} have developed rapidly in recent years and are widely used in natural language processing~\cite{wu2021deep}, financial analysis~\cite{wang2021review}, traffic prediction~\cite{diehl2019graph} or other fields. It can capture high-dimensional information between nodes by passing and aggregating messages on the topological structure of the graph. 

\textbf{Privacy concerns.} With the development of GNN, more and more graph datasets~\cite{hu2020open} are made public for researches and applications. Compared with other type of data, publishing graphs usually faces more severe privacy risks~\cite{han2022large}, since the graph includes not only node features, but also complex topology (edges). For example, in a social network, besides personal profiles, everyone's friend list also belongs to personal privacy that needs to be protected. However, Most of the previous work has ensured the features and labels privacy of nodes, while been not enough for edge privacy. In this paper, we consider the edge privacy and summarize the privacy protection scenarios as: upstream data owner publishes graph data for downstream data users' GNN tasks, but it needs to protect the edge privacy in the dataset before publishing.

\textbf{Challenge.} The biggest challenge in this graph privacy protection scenario is to make the published privacy preserving graph as usable as before. Specifically, edge information is part of graph data which is available to everyone who access the graph data. To preserve the edge privacy, there has to be a mechanism to perturb the edges in the graph, i.e. delete some real edges and add some false edges. However, GNN models use message passing strategy, which updates each node’s feature by aggregating its adjacent nodes. Therefore, the perturbing of edges may lead to bad information aggregation, for which privacy preservation of graph tends to negatively affect the performance or make the graph unavailable for downstream tasks. For instance, differential privacy (DP)~\cite{mueller2022sok}, a mainstream privacy protection method, protects edges by flipping 0s and 1s in the adjacency matrix with a certain probability. It will greatly reduce the accuracy of the model since some irrelevant nodes are connected by the added edges and some important edges which are hubs may be deleted.

\textbf{Prior work.} Some recent work have attempted to improve the performance of DP in GNN tasks, but their methods are usually hard to trade-off between the effect of privacy protection and the availability of data. For example, LapGraph~\cite{wu2022linkteller} adds Laplacian noise to the adjacency matrix to achieve perturbation; DPRR~\cite{hidano2022degree} adds and deletes edges by random sampling. However, both of them have a negative impact on information aggregation, making the model accuracy on the published graph much lower than that on the original graph. To solve the problem, \cite{lin2022towards} denoise the graph~ (modify the adjacency matrix to enhance graph smoothness) after DP, thus to improve the model accuracy. However, it weakens the protection effect because denoising process may make the published graph closer to the original graph.


\textbf{Our contributions.} In this paper, we design, apply, and evaluate a new framework for privacy graph publication, GraphPub, which is able to generate differential privacy graph with high availability. Our work is based on the assumption that for a graph, different edges are of different importance: Deleting or adding some particular edges may have little effect on the whole graph's information aggregation process. In other words, by delicately selecting the perturbed edges, the published privacy-preserving graph can achieve high performance on downstream GNN models. Such assumption is reasonable, because many works on explainable GNNs \cite{yuan2022explainability} \cite{cucala2021explainable} \cite{ying2019gnnexplainer} have already proved that: after extracting the main topology of graph and deleting all other edges, training on the modified graph composed of only a small number of edges can still ensure high task performance.

Based on this assumption, how to find these edges that have little influence on information aggregation becomes the critical problem. We divide the process of finding these edges into three steps: reverse learning, encoder-decoder and sampling. Specifically, use the original graph to train a GNN model $M$, and then learn a suppositional adjacency matrix $A_{s}$ through \textbf{reverse learning}~\cite{peng2022reverse} from $M$. According to the nature of reverse learning, $A_{s}$ will achieve the similar downstream performance compared with $A$. Then, to further generalize the topological information learned in $A_{s}$, we feed $A_{s}$ into a set of \textbf{encoder and decoder}~\cite{salehi2019graph}. The decoder outputs a probability matrix $L$. The larger $L_{i,j}$ is, the more likely there is an edge between node i and node j. Finally, according to the values in $L$, \textbf{sample} a certain number (depending on the requirements of privacy protection, i.e. privacy budget) of real edges and false edges to generate the protected adjacency matrix $\widetilde{A}$ for publication. Sufficient experiments prove that our edge protection method can achieve a high model accuracy when the privacy budget is extremely small.

We highlight our contributions as follows:
\begin{itemize}
\item For the first time, we apply reverse learning and decoder-encoder mechanism to DP, which is capable of generating highly available DP graph.

\item By combining DP and sampling, we protect the degree privacy of each node while ensuring that the degree distribution is basically unchanged, which makes the published graph have a similar sparsity to the original graph.

\item Our model maintains a high accuracy when privacy protection requirement is extremely strict (the privacy budget $\epsilon$ is very small, $\epsilon=1$).

\item Our model has good scalability and can be easily applied to GNN models like GCN, GAT, GraphSAGE. 

\item Our model can effectively defend against attackers for privacy data restoration like real edge prediction.
\end{itemize}

\section{Related Work}
\subsection{Differential Privacy}
Differential privacy is defined as follows~\cite{dwork2008differential}: A randomized algorithm $\mathcal{M}$ that takes a dataset consisting of individuals as input is ($\epsilon$, $\delta $)-differential private if for any pair of neighbouring data $D$, $D'$ that only differ in a single entry, and any result of this randomized algorithm y, and if $\delta$ = 0, we say that $\mathcal{M}$ is $\epsilon$-differential private.

\begin{equation}
Pr\left [ \mathcal{M}\left ( D \right )  =y \right ] \le e^{\epsilon} Pr\left [ \mathcal{M}  \left ( D' \right )=y \right ] +\delta 
\end{equation}

Most random algorithms which meet the requirements of differential privacy add noise that conforms to a certain distribution to the original data, making processed data indistinguishable.

Common differential privacy algorithms include Laplace Mechanism, Exponential Mechanism, Gaussian Mechanism, etc. Take the Laplace mechanism~\cite{igamberdiev2021privacy}\cite{qiu2022privacy} as an example. For a dataset $D$, the random query $Q$ obtains the sum of rows that meet a certain requirement, which is denoted as $h$, and then adds noise $x$ to the result $h$ to obtain a new result $y$. The noise $x$ satisfies the Laplace probability distribution with mean 0:

\begin{equation}
f\left ( x|\mu=0 ,\lambda \right )=\frac{1}{2\lambda}
e^{ -\frac{|x-\mu|}{\lambda}}
\end{equation}
where $\mu$ is the mean of the noise, and $\lambda$ is an undetermined parameter. 

Define global sensitivity $\bigtriangleup _{f}$ as Equation \ref{con:sensitivity}, which reflects the differences between neighboring datasets. 

\begin{equation}
\bigtriangleup _{f}=max_{D,D'}\left\{|h-h'|\right\}
\label{con:sensitivity}
\end{equation}

Let $\lambda=\frac{\bigtriangleup_{f}}{\epsilon }$, then the following inequality can be obtained, which shows Laplace mechanism is a $\epsilon$-differential private algorithm.

\begin{equation}
\begin{aligned}
\frac{Pr\left ( Q\left ( D \right ) = y \right ) }{Pr\left ( Q\left ( D' \right ) = y \right ) } 
&= \frac{Pr\left ( y-h \right ) }{Pr\left ( y-h' \right ) }\\
&=\frac{e^{-\frac{|y-h|}{\lambda } }}{e^{-\frac{|y-h'|}{\lambda } }} \\
&=e^{\frac{|y-h'|-|y-h|}{\lambda } }\\
&\le e^{\frac{|h-h'|}{\lambda } }\\
&=e^{\frac{|h-h'|}{\frac{\bigtriangleup_{f}}{\epsilon } } } \\
&\le e^{\epsilon } 
\end{aligned}
\label{con:Laplacian}
\end{equation}

However, the Laplacian mechanism cannot be used for data with binary attributes. A good example is an adjacency matrix, which consists of only 0s and 1s. For this binary data, random response mechanism~\cite{hidano2022degree}\cite{joshiedge} is used to protect its privacy. Given a privacy budget, we invert the original data with a certain probability $p$, as shown in Equation \ref{con:rr}.

\begin{equation}
\widetilde{a}_{i,j}=\left\{\begin{matrix}
a_{i,j}, &q=\frac{e^{\epsilon}}{1+e^{\epsilon}}  \\
1-a_{i,j},&p=\frac{1}{1+e^{\epsilon}}
\end{matrix}\right. 
\label{con:rr}
\end{equation}
where $a_{i,j}$ represents the elements in the adjacency matrix, $p$ and $q$ are the probabilities of inversion and keeping unchanged, respectively.

\subsection{Edge Privacy Protection}
Due to the binary nature of the adjacency matrix, random response mechanisms are widely used for edge privacy protection. Some papers have made improvements on the basis of DP, in order to ensure the availability of data after DP as much as possible. Lin et al.~\cite{lin2022towards} noticed that protecting the adjacency matrix through the random response mechanism would cause the graph to become dense, so they denoised the graph structure after DP to ensure the sparsity of the graph. However, denoising makes the disturbed graph close to the original graph again, which will undoubtedly reduce the effect of DP protection.

Hidano et al.~\cite{hidano2022degree} ensure that the degree distribution of the graph remains unchanged by randomly sampling after DP. This method is very simple, but because the sampling process is random, it may cause large changes in graph topology. Therefore, the model accuracy will drop significantly when the privacy budget is small.

Joshi et al.\cite{joshiedge} keep a certain number of real neighbors for each node, and then interfere with other neighbors through random response mechanism. This method does not strictly meet the requirements of DP, because when more real neighbors are preserved, it will lead to a large amount of privacy leakage. Therefore, even if it achieves high model accuracy, the effect of privacy protection cannot be guaranteed. 

Xu et al.\cite{xu2023mdp} achieve edge protection by employing matrix decomposition on the adjacency matrix. However, the computational complexity associated with matrix decomposition poses challenges for implementing this method when dealing with larger graph datasets.

As can be seen from the above methods, the main difficulty of edge privacy protection is to find a new adjacency matrix $\widetilde{A}$ in acceptable time. On the one hand, there is a large difference (depending on privacy budget) between $\widetilde{A}$ and the original adjacency matrix $A$, so as to protect the privacy data. On the other hand, $\widetilde{A}$ needs to ensure the rationality of the graph, so as to avoid a significant drop in model accuracy due to changes in topology.

\section{The Proposed GraphPub Framework}

\subsection{Security Model}

The data owner owns the original data of the graph, it will make the dataset public for downstream GNN tasks training such as academic research and commercial applications. However, since the data contains private information, it is necessary to protect the graph data before making it public. The requirements for processing of original graph have the following two parts: 

\begin{itemize}
\item Privacy protection. Realize the protection effect under a given privacy budget $\epsilon$, making it impossible for data users to determine whether any edge is real or false.
\item Data availability. The accuracy of the GNN model trained with the privacy-preserving graph should be as close as possible to the original graph.
\end{itemize}

This security model is very common in the real world. For example, Facebook's user information~\cite{ugander2011anatomy} is very important for studying social network models. However, since the user's friend list is personal privacy, Facebook must protect the privacy in the graph before making it public.

\subsection{GraphPub Overview}

\begin{figure*}[htbp]
\centerline
{\includegraphics[width=18cm]{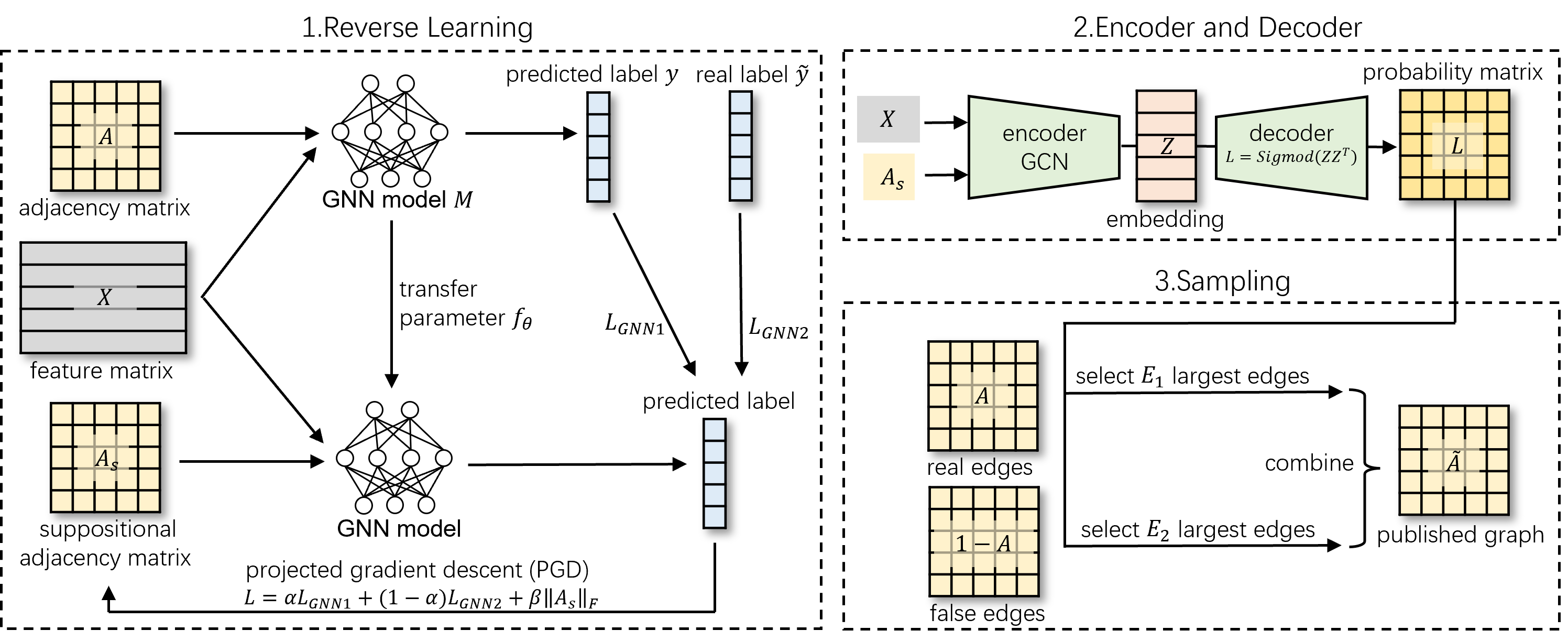}}
\caption{GraphPub Framework}
\label{fig: GraphPub framework}
\end{figure*}

Let the graph with $N$ nodes to be protected be $G = (A, X)$, where $A\in \mathbb{R} ^{N\times N}$ denotes the adjacency matrix, and $X\in \mathbb{R} ^{N\times F}$ denotes the features of the nodes. Use $G$ to train a GNN model, denoted as $M$. Through a specific loss function and gradient descent on $M$, we learn the suppositional adjacency matrix $A_{s}$. Due to the nature of reverse learning, the result of information aggregation on $A_{s}$ will be close to that on the original graph, that is, $A_{s}$ has a similar information aggregation effect to $A$. Then, use $A_{s}$ to train a two-layer GCN model for node classification. The first layer is used as an encoder~\cite{zhang2021graphmi} whose output is the hidden embedding of nodes, denoted as $Z\in \mathbb{R} ^{N\times H}$ ($H$ is the hidden layer dimension). Next, calculate the probability matrix by the decoder $L=sigmod(ZZ^{T})$ as a basis for sampling. Then according to privacy budget $\epsilon$, calculate the number of real and false edges ($E_{1}$ and $E_{2}$) owned by node i after DP. Finally, sample the $E_{1}$ real edges and $E_{2}$ false edges with the highest probability in $L$, and combine them to form a new adjacency matrix $\widetilde{A}$, which can be published and used for downstream GNN tasks. Figure \ref{fig: GraphPub framework} is the flow of the GraphPub framework.

This paper only deals with the protection of graph topology (edges). The protection of node features will be our future work.

\subsection{Reverse Learning}
Assume that the GNN model trained with the original graph is $M$, whose parameter is $f_{\theta}$. Initialize $A_{s}=0\times \mathbb{R} ^{N \times N}$. We optimize $A_{s}$ through the following loss function. 

\begin{equation}
\begin{split}
 Arg\ \underset{A_{s}}{min}& \ L = L_{GNN} +\beta \left \| A_{s} \right \|_{F}\\
L_{GNN} =& \alpha \frac{1}{N}  {\textstyle \sum_{i=1}^{N}} l_{i}(A_{s},f_{\theta },X,y_{i}) +\\
&(1-\alpha) \frac{1}{N} {\textstyle \sum_{i=1}^{N}} l_{i}(A_{s},f_{\theta },X,\widetilde{y}_{i})
\end{split}
\label{con: loss}
\end{equation}

The loss function consists of two parts. Among them, $L_{GNN}$ includes two parts: the loss of current model's output and $M$'s predicted label $y_{i}$, the loss of current model's output and the real label $\widetilde{y}_{i}$. Compare with the predicted label of $M$ to learn the input of M, i.e. the original adjacency matrix $A$; compare with the real label to promote the accuracy of the current model. 

Loss function's second term $\left \| A_{s} \right \|_{F}$ is the norm of $A_{s}$, in order to encourage the sparsity of the graph. 



In order to control the value range of $A_{s}$, after each gradient update, we use the projection function $P(\cdot)$ to map the elements in $A_{s}$ to the range of 0 to 1. This process is called projected gradient descent (PGD)~\cite{zhang2021graphmi}.

\begin{equation}
P(x)=\left\{\begin{matrix}
1,x>1 \\
0,x<0 \\
x,otherwise
\end{matrix}\right.
\end{equation}

\begin{equation}
A_s^{t+1}=P(A_s^t-\eta_t g_t)
\end{equation}
where $t$ is the number of iterations, $\eta$ is the learning rate, and $g$ is the gradient. 

The suppositional adjacency $A_{s}$ is obtained through PGD. Next, put it into a set of encoder and decoder to capture more general topological information.

\subsection{Encoder and Decoder}
Graph encoder and decoder provide us with a accurate representation (embedding) of node features. We train a two-layer node classification GCN with $A_s$ and $X$, whose first layer can be used as an encoder because it represents nodes' features with low-dimensional vectors. Furthermore, because GCN naturally uses graph topology for message passing and aggregation, these embedding is not only the extraction of node features, but also the extraction of graph topology. 

Multiplying encoder's output $Z$ and $Z^{T}$ gives a matrix whose elements are actually the dot products of nodes' embeddings. The larger the dot product between two nodes, the closer they are, the more likely there is an edge between them. Therefore, the probability matrix $L$ is obtained by the decoder showed as Equation \ref{con: L}.

\begin{equation}
L=sigmod(ZZ^{T})
\label{con: L}
\end{equation}

Both $L$ and $A_s$ are matrices of the same dimension as the adjacency matrix. But unlike $A_s$, each element of $L$ is a number between (0,1). Therefore, $L$ can more comprehensively reflect the possibility of an edge between any two nodes, thus providing more information for the sampling process.

\subsection{Sampling}
For a given privacy budget $\epsilon$, the differential privacy for edges is showed as Equation \ref{con:rr}. Suppose the degree of node i is $m_{i}$, that is, the i-th row of the adjacency matrix has $m_{i}$ 1s and $(N-m_{i})$ 0s. As shown in Equation \ref{con: m1m2}, node i has $m_{i}'$ edges after DP, including $m_{i,1}'$ real edges and $m_{i,2}'$ false edges. 

\begin{equation}
\left\{\begin{matrix}
m_{i}'=m_{i,1}'+m_{i,2}'\\
m_{i,1}'=\frac{e^{\epsilon}}{1+e^{\epsilon}} m_{i} \\
m_{i,2}'=\frac{1}{1+e^{\epsilon}} (N-m_{i})
\end{matrix}\right.
\label{con: m1m2}
\end{equation}

Due to the sparse nature of the graph ($N>>m$), DP makes graph become dense ($m_{i}'>>m_{i}$). To solve this problem, we restore the degree to $m_{i}$ by random sampling. For each edge, the probability of sampling is $m_{i}/m_{i}'$. After sampling, for the edge set of node i, there are $E_{1}$ real edges and $E_{2}$ false edges, as shown in Equation \ref{con: E1 and E2}. 

\begin{equation}
\left\{\begin{matrix}
 m_{i}=E_{i,1}+E_{i,2}\\
 E_{i,1}=\frac{m_{i}}{m_{i}'}  m_{i,1}'=\frac{e^{\epsilon} m_{i}}{e^{\epsilon} m_{i}+N-m_{i}} m_{i}\\
 E_{i,2}=\frac{m_{i}}{m_{i}'}  m_{i,2}'=\frac{N-m_{i}}{e^{\epsilon} m_{i}+N-m_{i}} m_{i}
\end{matrix}\right.
\label{con: E1 and E2}
\end{equation}

$E_{i,1}$ and $E_{i,2}$ are determined by the privacy budget $\epsilon$, the number of nodes $N$, and the degree $m_{i}$ of node i. When the privacy budget $\epsilon$ is smaller, $E_{i,1}$ is smaller and $E_{i,2}$ is larger, but their sum is always $m_{i}$. 

Combined with the decoder introduced above, $L$ reflects the rationality of the existence of an edge between two nodes. Therefore, when sampling $E_{i,1}$ real edges and $E_{i,2}$ false edges, choose the edges with higher values in $L$. Specifically, for node i, take out the i-th row of $L$, sort the values corresponding to the real edges, select the $E_{i,1}$ edges with the highest probability of existence, and add them to the new edge set. Similarly, select $E_{i,2}$ false edges with the highest probability to add to the new edge set. After selecting edges for each node, a DP adjacency matrix $\widetilde{A}$ that can be published is generated. 

\subsection{Degree Preservation}
In some cases, the degree of a node is also private information. Therefore, we need to use part of the privacy budget to protect the degree of nodes. The specific method is to divide the privacy budget $\epsilon$ into two parts $\epsilon_{1}$ and $\epsilon_{2}$, where $\epsilon_{1}$ is the privacy budget for node degree protection, and $\epsilon_{2}$ is the privacy budget for edge protection. We use the Laplacian mechanism to protect the degree of nodes, as shown in Equation \ref{con: Laplacian}. Finally, replace $m_{i}$ and $\epsilon$ in Equation \ref{con: E1 and E2} with $\widetilde{m}_{i}$ and $\epsilon_{2}$ respectively.

\begin{equation}
\widetilde{m}_{i} =max\left \{[m_{i}+Lap(\frac{1}{\epsilon _{1}} )], 1\right \} 
\label{con: Laplacian}
\end{equation}
where $Lap(\lambda)$ represents Laplacian noise with mean 0 and scale $\lambda$, $[\cdot]$ indicates rounding.

\section{Experiment}
In this section, we present the experimental results to show the effectiveness of GraphPub, we pose the following research questions:

\begin{itemize}
\item \textbf{RQ1.} How accurate is our GraphPub, and how does it compare to other algorithms?
\item \textbf{RQ2.} How much does our GraphPub preserve each user’s degree information?
\item \textbf{RQ3.} Can we do some ablation study to show that the different modules are actually effective?
\item \textbf{RQ4.} How compatible is our model?
\item \textbf{RQ5.} Can our model effectively defend against attacks?
\item \textbf{RQ6.} What is the overhead of our model?
\end{itemize}

Different experiments are designed to answer these questions. We will introduce the experimental setup and show experimental results in the following sections.

\subsection{Experimental Set-up}
\textbf{Dataset.} We conducted experiments over three datasets, which can be divided into two categories: citation networks and social networks.

\begin{itemize}
\item \emph{Citation Networks:} We used Cora and Citeseer~\cite{sen2008collective}. Among them, nodes represents words features and edges represent the citations among nodes.
\item \emph{Social Networks:} Polblogs~\cite{adamic2005political} dataset consists of information about U.S. political blogging sites, includes blogging sites and hyperlinks. Every blog has policical attributes: conservative or liberal.
\end{itemize}

\begin{table}[htbp]
	\centering
        \resizebox{6cm}{!}{
	\begin{tabular}{ccccc}
		\toprule 
		Datasets & Nodes & Edges & Classes & Features \\ 
  		\midrule
		Cora & 2708 & 5429 & 7 & 1433 \\
		Citeseer & 3327 & 4732 & 6 &  3703 \\
		Polblogs & 1490 & 19025 & 2 & - \\
		\bottomrule
	\end{tabular}
        }
        \caption{Summary of Datasets}
	\label{datasets}
\end{table}

\textbf{Test model.} In order to test the performance of the graph after GraphPub on different models. We use three popular GNN models: GCN~\cite{welling2016semi}, GAT~\cite{velickovic2017graph}, and GraphSAGE~\cite{hamilton2017inductive}. Model parameters are the same as those in the original literature. During the training of the model, 10\% nodes are randomly selected as training set. All GNN models are trained for 200 epochs by using an early stopping strategy based on convergence behavior and  accuracy on a verification set contains 20\% randomly selected nodes.

\textbf{Baselines.} There are three baseline methods, RR~\cite{lin2022towards}, DPRR~\cite{hidano2022degree} and LAPGRAPH~\cite{wu2022linkteller}:

\begin{itemize}
\item \textbf{RR.} RR is a neat differential privacy algorithm. It flips each 0/1 with probability $1/(e^{\epsilon}+1)$. It provides $\epsilon-edge$ (DP on edges with privacy budget $\epsilon$).
\item \textbf{DPRR.} DPRR is similar to RR, also providing $\epsilon-edge$, where $\epsilon=\epsilon_1+\epsilon_2$. But DPRR allocates a portion of the privacy budget $\epsilon_1$ to degree privacy protection and adds EdgeSampling module to solve the problem of DP graph being too dense in RR.
\item \textbf{LAPGRAPH.} LapGraph calculates the number of edges by applying the Laplacian mechanism to a small fraction of the privacy budget $\epsilon_1$, The remaining privacy budget $\epsilon_2=\epsilon-\epsilon_1$ is used to apply the Laplacian mechanism to the entire adjacency matrix. The top-elements in the perturbed adjacency matrix are set to 1 and the remaining elements are set to 0.
\end{itemize}

\textbf{Parameter settings.} In the experiments, we set $\alpha=0.5, \beta=0.0001, \eta_t=0.1$ in Equation \ref{con: loss} as default setting.

\subsection{Experimental Result}

\begin{table*}
    \centering
    \resizebox{18cm}{!}{
    \begin{tabular}{c|clp{2cm}p{2cm}p{2cm}p{2cm}p{2cm}p{2cm}p{2cm}p{2cm}}
        \toprule
        Datasets  & Models & $\epsilon$ & 20 & 17 & 14 & 11 & 8 & 5	& 2 & 1 \\
        \hline
        \multirow{12}*{Cora} & \multirow{4}*{GCN} & RR & \textbf{0.8002} & 0.795 & 0.8004 & 0.8002 & 0.7298 & 0.3984 & 0.3098 & 0.3074\\
        & & DPRR & 0.794 & 0.7848 & 0.7186 & 0.4408 & 0.2402 & 0.1804 & 0.18 & 0.1674\\
        & & LAPGRAPH & 0.7898 & 0.7516 & 0.6896 & 0.4658 & 0.3106 & 0.3114 & 0.3116 & 0.2794\\
        & & GraphPub & 0.7996 & \textbf{0.7962} & \textbf{0.8162} & \textbf{0.8242} & \textbf{0.8202} & \textbf{0.811} & \textbf{0.8064} & \textbf{0.7806}\\
        \cline{2-11}
        & \multirow{4}*{GAT} & RR & \textbf{0.7918} & \textbf{0.7884} & 0.7896 & 0.7874 & 0.712 & 0.3798 & 0.0944 & 0.2012\\
        & & DPRR & 0.7838 & 0.7732 & 0.6978 & 0.4422 & 0.2146 & 0.1632 & 0.1542 & 0.1736\\
        & & LAPGRAPH & 0.7702 & 0.7458 & 0.6732 & 0.4556 & 0.1012 & 0.1702 & 0.1156 & 0.1452\\
        & & GraphPub & 0.7872 & 0.787 & \textbf{0.8038} & \textbf{0.8228} & \textbf{0.8114} & \textbf{0.8024} & \textbf{0.7926} & \textbf{0.7644}\\
        \cline{2-11}
        & \multirow{4}*{GraphSAGE} & RR & 0.7656 & \textbf{0.7698} & 0.7714 & 0.7716 & 0.7228 & 0.5132 & 0.5414 & 0.5132\\
        & & DPRR & 0.7678 & 0.7624 & 0.7072 & 0.5378 & 0.3944 & 0.3228 & 0.3178 & 0.3162\\
        & & LAPGRAPH & 0.7602 & 0.739 & 0.693 & 0.5618 & 0.5118 & 0.532 & 0.5442 & 0.5176\\
        & & GraphPub & \textbf{0.7682} & 0.7654 & \textbf{0.7816} & \textbf{0.7934} & \textbf{0.7904} & \textbf{0.7738} & \textbf{0.778} & \textbf{0.7484}\\
        \hline
        \multirow{12}*{Citeseer} & \multirow{4}*{GCN} & RR & \textbf{0.656} & 0.6508 & 0.6566 & 0.641 & 0.5298 & 0.5298 & 0.187 & 0.2002\\
        & & DPRR & 0.6548 & 0.6332 & 0.5434 & 0.3386 & 0.2352 & 0.2024 & 0.2122 & 0.2076\\
        & & LAPGRAPH & 0.6338 & 0.5954 & 0.525 & 0.3188 & 0.2094 & 0.1958 & 0.1952 & 0.1982\\
        & & GraphPub & 0.6556 & \textbf{0.6558} & \textbf{0.689} & \textbf{0.7092} & \textbf{0.7256} & \textbf{0.717} & \textbf{0.6884} & \textbf{0.6742}\\
        \cline{2-11}
        & \multirow{4}*{GAT} & RR & \textbf{0.6372} & 0.6366 & 0.6382 & 0.6352 & 0.5104 & 0.5104 & 0.1596 & 0.1642\\
        & & DPRR & 0.6314 & 0.6172 & 0.5274 & 0.3274 & 0.2312 & 0.2122 & 0.213 & 0.2126\\
        & & LAPGRAPH & 0.6174 & 0.5792 & 0.5084 & 0.3154 & 0.1762 & 0.1854 & 0.2002 & 0.1852\\
        & & GraphPub & 0.637 & \textbf{0.6446} & \textbf{0.6774} & \textbf{0.6998} & \textbf{0.714} & \textbf{0.7016} & \textbf{0.6786} & \textbf{0.6754}\\
        \cline{2-11}
        & \multirow{4}*{GraphSAGE} & RR & 0.6434 & 0.6434 & 0.6448 & 0.6524 & 0.5738 & 0.449 & 0.495 & 0.5168\\
        & & DPRR & 0.6444 & 0.6438 & 0.5824 & 0.457 & 0.3634 & 0.347 & 0.3458 & 0.324\\
        & & LAPGRAPH & 0.6242 & 0.609 & 0.5642 & 0.421 & 0.4696 & 0.4942 & 0.5062 & 0.5182\\
        & & GraphPub & \textbf{0.6456} & \textbf{0.6504} & \textbf{0.6726} & \textbf{0.695} & \textbf{0.6996} & \textbf{0.6846} & \textbf{0.6698} & \textbf{0.6626}\\
        \hline
        \multirow{12}*{Polblogs} & \multirow{4}*{GCN} & RR & 0.8488 & 0.8543 & 0.8543 & \textbf{0.8638} & \textbf{0.8482} & \textbf{0.7916} & 0.5728 & 0.4973\\
        & & DPRR & 0.8486 & 0.8552 & 0.8379 & 0.8057 & 0.6838 & 0.5099 & 0.5105 & 0.5042\\
        & & LAPGRAPH & \textbf{0.8512} & 0.8587 & 0.8274 & 0.7973 & 0.7638 & 0.5011 & 0.5021 & 0.4992\\
        & & GraphPub & 0.8478 & \textbf{0.8752} & \textbf{0.8579} & 0.8331 & 0.8434 & 0.7234 & \textbf{0.6943} & \textbf{0.6693}\\
        \cline{2-11}
        & \multirow{4}*{GAT} & RR & 0.8712 & 0.868 & \textbf{0.8688} & \textbf{0.8701} & \textbf{0.8556} & 0.8008 & 0.5895 & 0.4977\\
        & & DPRR & \textbf{0.8749} & \textbf{0.8695} & 0.8615 & 0.8162 & 0.6905 & 0.5236 & 0.5055 & 0.5076\\
        & & LAPGRAPH & 0.8663 & 0.8573 & 0.8278 & 0.808 & 0.7669 & 0.488 & 0.5078 & 0.4922\\
        & & GraphPub & 0.8714 & 0.8667 & 0.8686 & 0.8638 & 0.8512 & \textbf{0.816} & \textbf{0.7968} & \textbf{0.8305}\\
        \cline{2-11}
        & \multirow{4}*{GraphSAGE} & RR & 0.8596 & \textbf{0.8627} & \textbf{0.8617} & 0.856 & \textbf{0.8497} & 0.7389 & 0.572 & 0.4981\\
        & & DPRR & \textbf{0.8621} & 0.8621 & 0.8518 & 0.8034 & 0.6863 & 0.5183 & 0.4964 & 0.4958\\
        & & LAPGRAPH & 0.8472 & 0.8482 & 0.8173 & 0.7851 & 0.7097 & 0.4901 & 0.5034 & 0.5015\\
        & & GraphPub & 0.8606 & 0.8623 & 0.8596 & \textbf{0.8585} & 0.8484 & \textbf{0.8307} & \textbf{0.7874} & \textbf{0.8324}\\
        \bottomrule
    \end{tabular}
    }
    \caption{Model Accuracy}
    \label{tab:Accuracy}
\end{table*}

\begin{figure*}[!ht]
    \centering
    \subfloat[Degree Similarity]{\includegraphics[width=4.2cm]{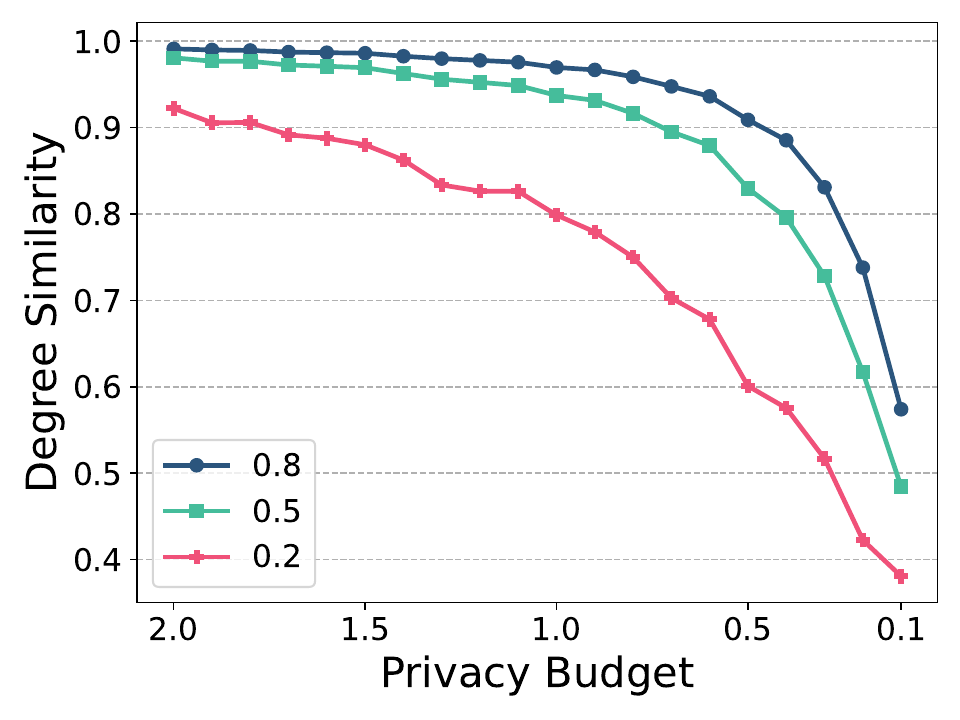} 
    \label{fig: Degree Similarity}}
    \subfloat[Degree Preservation]{\includegraphics[width=4.2cm]{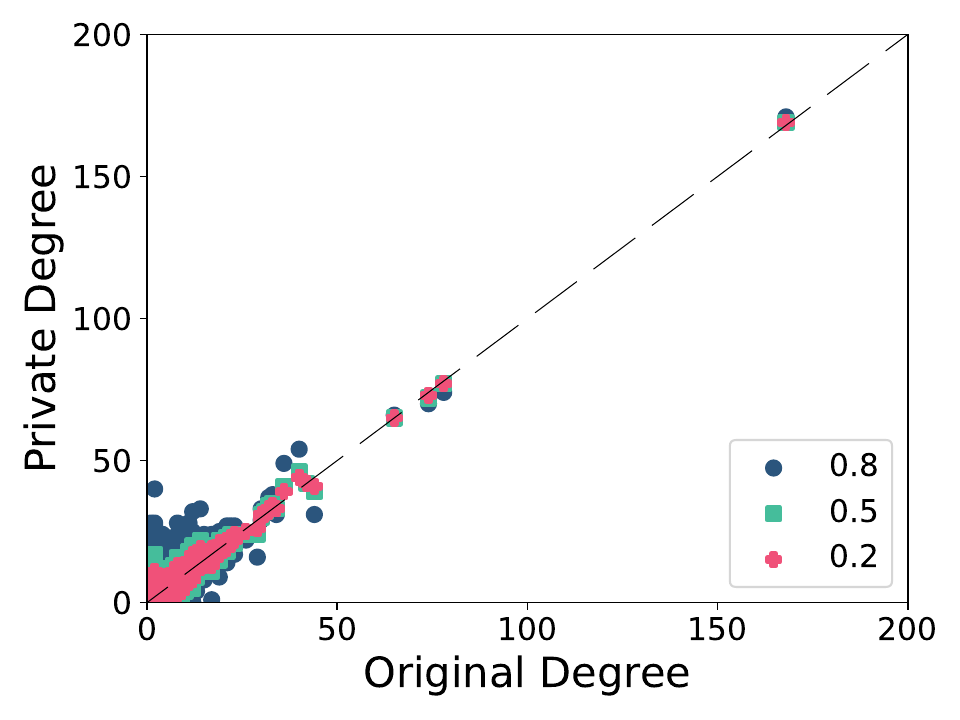} 
    \label{fig: Degree Preservation}}
    \subfloat[Probability Matrix Ablation]{\includegraphics[width=4.2cm]{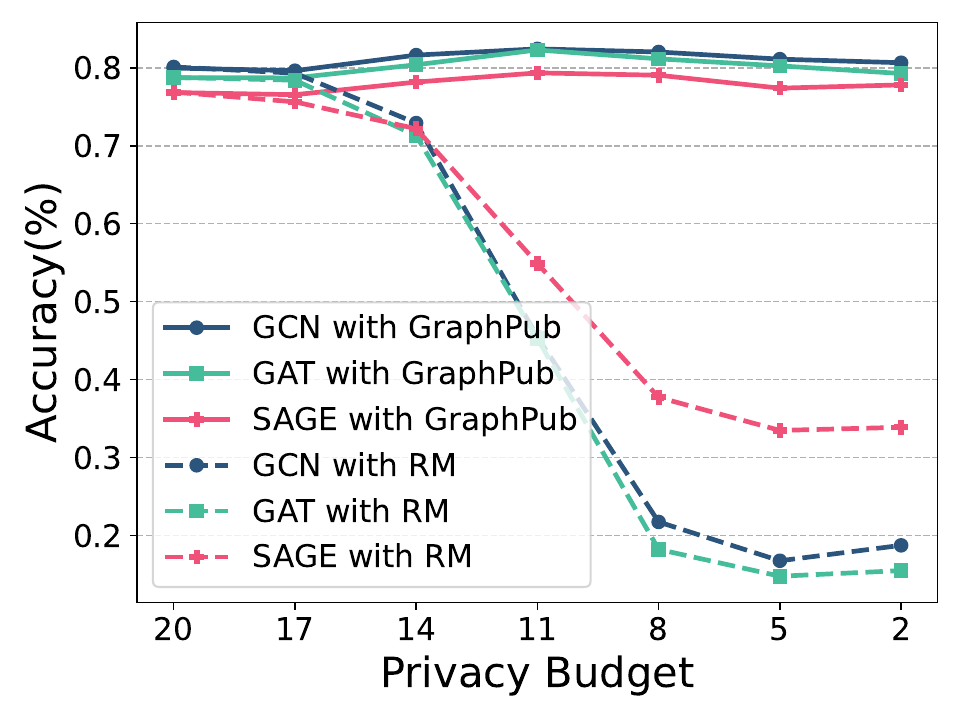} 
    \label{fig: Probability Matrix Ablation Study}}
    \subfloat[PGD Ablation]{\includegraphics[width=4.2cm]{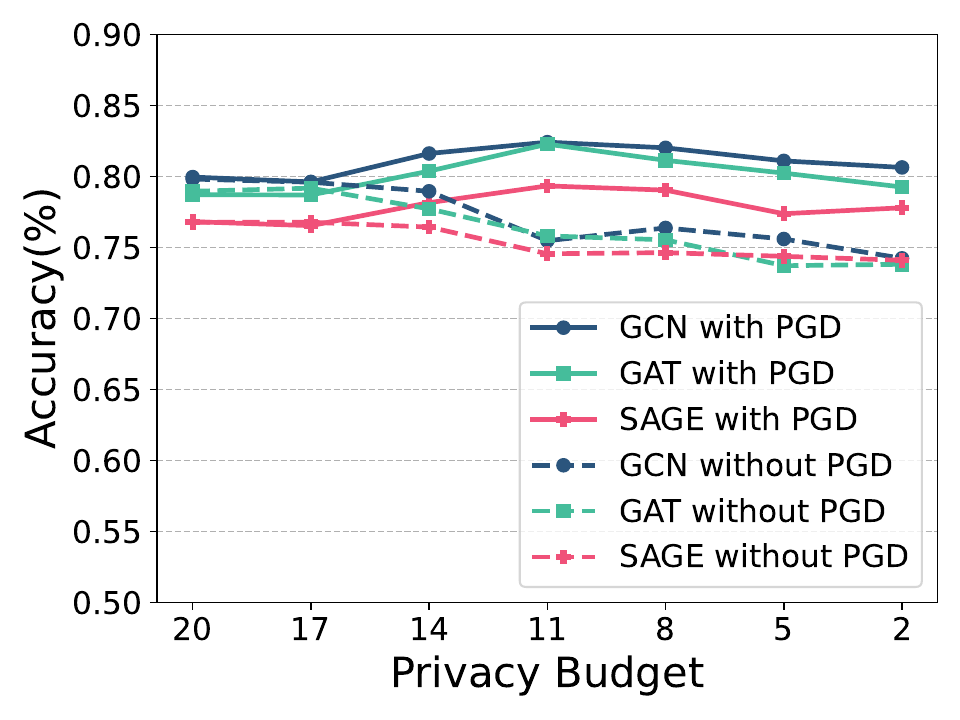} 
    \label{fig: PGD Ablation Study}}
    \caption{Experimental Results}
    \label{fig: results}
\end{figure*}

\textbf{RQ1. Accuracy.} We set three baselines (RR, DPRR and LAPGRAPH) as comparison, set the privacy budget between 20 and 0, and sample the important parts as Table \ref{tab:Accuracy}. We use three GNN models, namely GCN, GAT and GraphSAGE, to test the classification accuracy of the DP graph. The experiment was carried out on three datasets, and the experimental results were the mean values after five repeated experiments to ensure that there would be no deviation.

Looking at our GraphPub alone, we can see that with the decrease of privacy budget, the accuracy of GraphPub does not lose too much, which shows the superiority of our method. At the same time, we also observed an increase in accuracy as privacy budgets declined, which is reasonable, Because GraphPub will increase the importance of some edges that play a positive role in classification accuracy in the process of gradient descent, which is reflected in the increase of the probability of these edges in the final probability matrix $L$. In addition, it can be seen that although our method is trained on the GCN model, it performs well in the tasks of GCN, GAT and GraphSAGE. This shows that our DP graph is universal.

Observe our GraphPub and RR,DPRR, LAP GRAPH. It can be seen that in almost all cases of privacy budget, our method achieves no less accuracy than the Baselines methods, which shows that our method performs better than the baseline methods in classification accuracy experiments.

\textbf{RQ2. Degree Preservation.} Next, we will explore the performance of GraphPub on Degree Preservation. Since the degree of a user has a very small sensitivity ~\cite{hidano2022degree}, Large privacy budgets may cause a very small Laplacian perturbation, we set the privacy budget $\epsilon$ between 2 and 0. In the experiment, We allocate 0.8, 0.5, 0.2 of the total privacy budget $\epsilon$ to the degree-preserving privacy budget $\epsilon_1$ and conduct experiments on the Cora dataset. As a judgment criteria, we count the degree of each node in the original graph and the DP graph respectively, and use cosine similarity to calculate the degree similarity between them. The results are shown in Figure \ref{fig: Degree Similarity} . In order to reflect the degree distribution more intuitively, we show the degree distribution with privacy budget $\epsilon=1$ in Figure \ref{fig: Degree Preservation}.



It can be seen from the experiment that GraphPub can retain degree information well, and the retention of degree information can be controlled by adjusting the proportion of the budget allocated to degree-preserving privacy. In a practical application, the proportion of the privacy budget can be set according to whether the retention degree information is required.

\textbf{RQ3. Ablation Study.} In order to verify the actual role of each algorithm in GraphPub, we designed two experiments, the first is to replace the probability matrix $L$ with a random matrix (RM), and the second is to remove the PGD module, we put the original graph into the Graph Autoencoder. The experiment is carried out on the Cora dataset, and the privacy budget is from 20 to 0. The result is the accuracy of the test set after the DP graph is retrained on the GCN.

The result of the first experiment is shown in Figure \ref{fig: Probability Matrix Ablation Study}. It can be seen that when the probability matrix is replaced by a random matrix, the accuracy rate will drop rapidly, which is shown in the three training models, which shows the effectiveness of our probability matrix.


The result of the second experiment is shown in Figure \ref{fig: PGD Ablation Study}. It can be seen that after the PGD module is removed, the classification accuracy also decreases, which shows that our gradient descent module plays a positive role in preserving the classification accuracy.


It also can be seen from Figure \ref{fig: PGD Ablation Study} that GraphPub still has good classification accuracy even if the gradient descent part is removed. Gradient descent is a module that consumes a lot of time. If the task has higher requirements for time and relatively low requirements for accuracy, it can be considered to remove the gradient descent part to improve the speed of the model.

\begin{figure}[htbp]
\centerline
{\includegraphics[width=5cm]{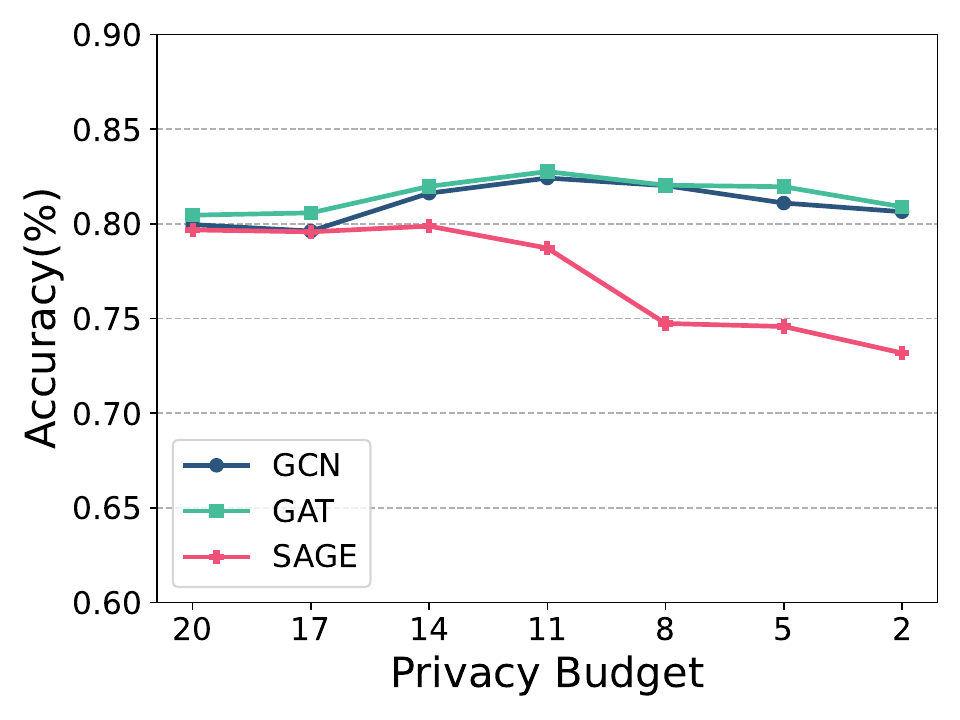}}
\caption{Extendibility}
\label{fig: Extendibility}
\end{figure}

\textbf{RQ4. Extendibility.} Note that we use the GCN model to construct the probability matrix. For the GraphPub model, we support a variety of models to construct the probability matrix. We designed the training of GraphPub model on GAT and SAGE respectively, and tested the classification accuracy of the test set by re-training the DP graph on GCN on the Cora dataset. The results are shown in Figure \ref{fig: Extendibility}.

Under different models, GraphPub can still maintain a high classification accuracy, which shows that GraphPub has good scalability. If there is a better GNN model in the future, GraphPub can easily replace to the latest model.

\textbf{RQ5. Attack Test.} Since our published graph contains graph features and graph topology, the relationship between A and B, B and C may reveal the relationship between A and C through Embedding Similarity Attack. 

\emph{Embedding Similarity Attack: It is based on the assumption that nodes with more similar embeddings are more likely to be adjacent. Therefore, the attacker uses the cosine similarity of node embeddings to infer edges. The node embedding can be obtained by the node representation of the hidden layer of the GNN model.}

We conduct experiments on Cora, Citeseer and Polblogs, set the privacy budget to 1, and use Embedding Similarity to attack GraphPub. We assume that the attacker can somehow estimate a conjectured number of edges $\widetilde{E}$ that approximates the number of edges in the original graph. This is a reasonable assumption and reduces the difficulty of the attack. The attacker will select the first n edges that are most likely to exist according to the Embedding Similarity and speculate them as real edges. We examine the precision and recall values of these edges relative to the real edges to determine whether the attacker can succeed.

The results of the experiment are shown in Table \ref{tab:Precision and Recall}. It can be seen that precision and recall are very low, indicating that the results of the attack are of little value to the attacker. Considering that we do not protect node features, while the features will improve the effect of the attack, we believe that this result is enough to show that the attacker still can not attack GraphPub effectively under a relatively tolerant condition.

\begin{table}[htbp]
    \centering
    \resizebox{5.5cm}{!}{
    \begin{tabular}{llcc}
    \toprule
     &    & Precision & Recall \\
    \hline
    \multirow{2}*{Cora} & $A_t$ vs $A$ & 0.0265 & 0.0402 \\
     & $A_t$ vs $\widetilde{A}$  & 0.1592 & 0.2221 \\
    \hline
    \multirow{2}*{Citeseer} & $A_t$ vs $A$ & 0.0284 & 0.0269 \\
     & $A_t$ vs $\widetilde{A}$  & 0.1927 & 0.1679 \\
    \hline
    \multirow{2}*{Polblogs} & $A_t$ vs $A$ & 0.0362 & 0.0369 \\
     & $A_t$ vs $\widetilde{A}$ & 0.1488 & 0.1487 \\
    \bottomrule
    \end{tabular}
    }
    \caption{Precision and Recall, $A_t$ represents the graph obtained by the attacker, while $A$ and $\widetilde{A}$ represent the original graph and the published DP graph, respectively.}
    \label{tab:Precision and Recall}
\end{table}

As a comparison, we show the precision and recall between the attack result and our DP graph. We can see that the precision and recall between attack result and our DP graph is about five times that of the original graph. The attack result using the published graph is closer to the published graph than the original graph, which prevents the original graph privacy from being restored.

\textbf{RQ6. Overhead.} We compared the running time of GraphPub with the training time of mainstream GNN models on RTX 2080Ti x2  with 128GB  memory, and the experimental results are shown in Table \ref{tab: Overhead}. The results prove that our model overhead is comparable to ordinary GNN model training, which is acceptable in privacy preserving scenario.

\begin{table}[htbp]
    \centering
    \resizebox{4.8cm}{!}{
    \begin{tabular}{cccc}
    \toprule
    GraphPub & GCN & GAT & GraphSAGE \\
    \hline
    1513 & 717 & 4514 & 1288 \\
    \bottomrule
    \end{tabular}
    }
    \caption{Overhead, dataset: Cora, unit: ms.}
    \label{tab: Overhead}
\end{table}

\section{Conclusion}
This paper focuses on the privacy issue of graph publishing containing sensitive data, that is, upstream data owners maintain graph availability while protecting privacy, and finally publish graph data for downstream GNN task. Differential privacy is the mainstream method for privacy protection, but random changes to the graph topology in DP will reduce the availability of graph data (specifically, the accuracy of the GNN model will decrease). We design a reverse learning and encoder-decoder framework (GraphPub) for better edge sampling in DP. Sufficient experiments show that our model can still maintain the model accuracy close to the original graph when the privacy budget is extremely small. Besides, our model also has high scalability and strong anti-attack capabilities. However, GraphPub requires the data owner to have certain computing power (able to train the GNN model). In addition, our work does not include the protection of node features. Therefore, we leave reducing the model overhead and providing node feature protection as our future work.

\clearpage

\bibliographystyle{named}
\bibliography{ijcai23}

\end{document}